\documentclass[prl,twocolumn,groupedaddress]{revtex4}

\usepackage{graphicx}
\usepackage{calc}
\usepackage{pifont}
\usepackage{amssymb}
\usepackage{epsfig}
\usepackage{amssymb}
\usepackage{amsmath}
\usepackage{color}
\usepackage{braket}

\usepackage{url}

\DeclareMathAlphabet{\mathitb}{OT1}{cmr}{bx}{sl}
\begin{document}

\renewcommand{\thefootnote}{\fnsymbol{footnote}}
\title{Transport Characterization of Kondo-Correlated Single Molecule Devices}
\author{G. D. Scott$^1$}
\email{gavin.scott@alcatel-lucent.com}
\author{D. Natelson$^{2,3}$}
\author{S. Kirchner$^{4,5}$}
\author{E. Mu\~{n}oz$^{6}$}
\affiliation{
$^{1}$Bell Laboratories, Alcatel-Lucent, 600 Mountain Ave, Murray Hill, NJ 07974\\
$^{2}$Department of Physics and Astronomy and $^{3}$Department of Computer and Electrical Engineering, Rice University, 6100 Main St, Houston, TX 77005\\
$^{4}$Max Planck Institute for the Physics of Complex Systems, 01187 Dresden, Germany\\
$^{5}$Max Planck Institute for Chemical Physics of Solids, 01187 Dresden, Germany\\
$^{6}$Facultad de F\'{i}sica, Pontificia Universidad Cat\'{o}lica de Chile, Casilla 306, Santiago 22, Chile}

\date{\today}

 \begin{abstract}

A single molecule break junction device serves as a tunable model system for probing the many body Kondo state. The low-energy properties of this state are commonly described in terms of a Kondo model, where the response of the system to different perturbations is characterized by a single emergent energy scale, $k_{\mathrm{B}}T_K$. Comparisons between different experimental systems have shown issues with numerical consistency. With a new constrained analysis examining the dependence of conductance on temperature, bias, and magnetic field simultaneously, we show that these deviations can be resolved by properly accounting for background, non-Kondo contributions to the conductance that are often neglected.  We clearly demonstrate the importance of these non-Kondo conduction channels by examining transport in devices with total conductances exceeding the theoretical maximum due to Kondo-assisted tunneling alone.


\end{abstract}

\maketitle

Strong interactions between charge carriers can lead to the emergence of novel many-body transport phenomena. In some cases the resulting states may exhibit universal characteristics that bridge seemingly diverse systems.  The Kondo effect is an archetype of such correlated electron behavior, in which a collective state is established when itinerant conduction electrons antiferromagnetically screen a local magnetic moment.

Innovative device configurations possessing a small island tunnel-coupled to a Fermi sea allow for the formation of a localized magnetic impurity, represented here as a single unpaired spin.  Examples include semiconductor quantum dots,\cite{Goldhaber1998a,Cronenwett1998} nanowires,\cite{Kretinin2011} carbon nanotubes,\cite{Nygard2000,Herrero2005}, individual molecules,\cite{JPark2002,Liang2002} including C$_{60}$,\cite{Yu2004,Pasupathy2004,Parks2007} and in STM measurements of both magnetic adatoms on metallic surfaces\cite{Madhavan1998} and in chemically homogenous magnetic materials~\cite{Calvo2009}.  Kondo screening occurs when the unpaired spin forms a singlet state with the conduction electrons as they move on and off the island via elastic cotunneling processes.  This results in the quenching of the local spin, and produces an extra peak in the density of states at the Fermi level (Figs. \ref{figure1}a and b). The hallmark signature of the fully screened spin-$\frac{1}{2}$ Kondo effect in these structures is an enhanced zero-bias conductance that forms below an emergent energy scale known as the Kondo temperature, $T_K$ ($k_{\mathrm{B}}=1$), which represents the binding energy of the many-body singlet state.

The spin-$\frac{1}{2}$ Kondo model describes the antiferromagnetic interaction between the magnetic moment and conduction electrons at strong coupling in a particle-hole symmetric device~\cite{Schiller1998,Majumdar1998,Kretinin2011}.  A remarkable feature of this model is the formation of a local Fermi liquid state characterized by a single energy scale, $T_K$.  As a result, the low-energy behavior is expected to be universal.  Realistic devices can deviate from this expectation for a number of reasons.  This could occur, for example, due to a departure from particle-hole symmetry\cite{Munoz2013} or due to multiple energy levels on the island participating in transport.  To fully characterize the electronic properties of devices, a reliable extraction of $T_K$, and additional parameters, is critical to the ensuing analysis.  Previous analyses for $T_K$ and a number of transport coefficients, based on semi-phenomenological formulas for the spin-$\frac{1}{2}$ Kondo model, have found agreement between different gated semiconductor systems, but systematic disagreement with single molecule devices.\cite{Grobis2008,Kretinin2011,Scott2009}.

In this paper we report an analysis of conduction measurements through single molecule devices in the low energy regime as a function of temperature, bias, and magnetic field. We observe zero-bias resonance peaks with total conductance, $G$, exceeding the unitary limit ($G_0 > 2e^2/h$), which serves to highlight the presence of extra transport channels in addition to Kondo scattering.  Our analysis allows us to characterize the low-energy behavior of these devices in a consistent manner in terms of an Anderson model and a background contribution.  Most investigations ignore non-Kondo conductance channels\cite{Goldhaber1998b,Grobis2008}, as they are assumed to be vanishingly small or simply irrelevant.  We demonstrate that this background contribution can be non-negligible in single molecule devices fabricated with a break junction technique. Properly separating Kondo and non-Kondo transport is essential in obtaining consistent values for $T_K$ and the corresponding transport coefficients, and may impact our understanding of the effective low-energy model of real devices.

Fits to the frequently employed phenomenological equation for $G(T)$, together with recently developed analytical descriptions for $G(V)$ and $G(B)$ in the spin-$\frac{1}{2}$ Kondo model (see below), provide three independent means of inferring the characteristic energy scale $T_K$.  We determine a background conductance offset by including it as a free fitting parameter for the three fits, assuming that it is weakly dependent upon temperature, bias voltage and magnetic fields for energies $< k_{\mathrm{B}}T_K$ and may thus be modeled as a constant.  The ``optimum'' offset due to background conduction is the one that most nearly produces the same $T_K$ for the different perturbations.  This treatment is applied to both new data and to previously published results, potentially resolving the conflicting coefficient values found in different systems for the ostensibly universal transport coefficients.

The metallic break junction devices we used were defined by e-beam lithography on $n^+$ Si substrate with 200~nm SiO$_2$.  They were composed of 15~nm Pd, deposited by ebeam evaporation.  The resulting constriction patterns possessed minimum widths between 40~nm and 80~nm.  Devices were cleaned first with a standard solvent rinse followed by an oxygen plasma to remove residual organic compounds.  A single drop of 100~${\mu}M$ C$_{60}$ in toluene solution was placed on an array of devices for 30 sec and then blown off with N$_2$ gas.  Measurements were performed inside of a $^3$He-$^4$He dilution refrigerator with a base temperature $<~100~mK$. Thermally-assisted electromigration was employed at temperatures below 2~K to create a tunneling gap.\cite{Park1999}  Differential conductance ($dI/dV$) was measured as a function of source-drain bias ($V$) using standard low frequency lock-in techniques (Fig. \ref{figure1}c).  Identification of devices possessing the desired electrode-C$_{60}$-electrode geometry is discerned from the resulting transport characteristics, as has been discussed elsewhere~\cite{Scott2010b,Yu2004}

The Kondo ground state is expected to be a Fermi liquid, and thus the lowest order correction to $G(T=0,V=0,B=0)$ is quadratic in temperature, bias voltage and magnetic field. When normalized by $T_K$, the response of the system to different perturbations can be described by equations of the same form at low energies ($k_{\mathrm{B}}T,eV,|g|{\mu_B}B \ll k_{\mathrm{B}}T_K$),
\par\nobreak
\vspace{-4mm}
\footnotesize
\begin{eqnarray}
G(T) \equiv G(T,0,0) &=& G(0,0,0) - c_T G_0(T/T_K)^2, \\
G(V) \equiv G(0,V,0) &=& G(0,0,0) - c_V G_0(eV_{sd}/k_{\mathrm{B}}T_K)^2,\\
G(B) \equiv G(0,0,B) &=& G(0,0,0) - c_B G_0 (|g|{\mu_B}B/k_{\mathrm{B}}T_K)^2,
\label{Fermi}
\end{eqnarray}
\normalsize
where $G_0 \equiv G(0,0,0)-G_b$ with $G_b$ being the background contribution and $c_T (c_V, c_B)$ is a transport coefficient in $T$ ($V, B$), respectively.  The equilibrium conductance in the Kondo regime measured in a variety of systems~\cite{Goldhaber1998a,Nygard2000,Yu2004} has been shown to obey a universal temperature dependence described by the following phenomenological expression\cite{Goldhaber1998b} that smoothly connects the low ($T/T_K \ll 1$) and intermediate energy ($T/T_K \sim 1$) regimes,
\begin{equation}
G(T) = G_0\left[1 + \left(2^{1/s} - 1\right)\left(\frac{T}{T_K}\right)^{2} \right]^{-s}+ G_b,
\label{GEK}
\end{equation}
where $s = 0.21$ for a spin-$\frac{1}{2}$ Kondo model in the absence of any potential scatterer.\cite{Costi1994,Goldhaber1998b} The Kondo temperature is defined by $[G(T = T_K) - G_b]/G_0 = 1/2$.  Since $k_{\mathrm{B}}T_K$ is a crossover energy scale rather than a sharp transition, there is some ambiguity in its definition.  However, numerical agreement should be expected between analyses based on a consistent definition of $T_{K}$.  Eq. (\ref{GEK}) has been widely used as the principle means of estimating $T_K$ in experimental systems, although $G_b$ is often not included.

\begin{figure}[t!]
\begin{center}
\includegraphics[scale = .2]{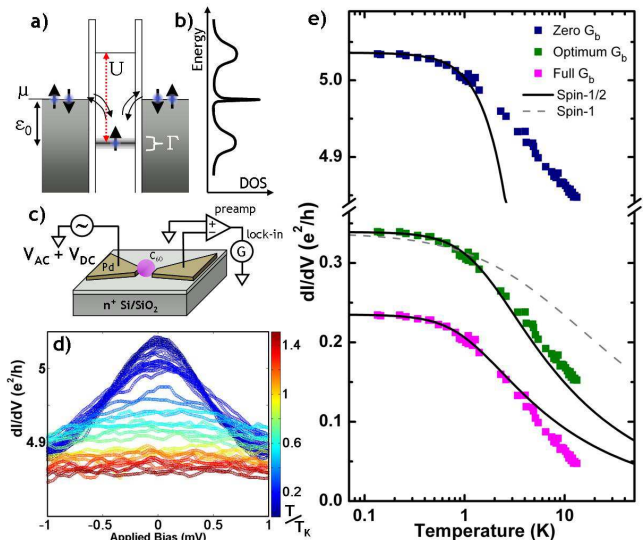}
\end{center}
\vspace{-7mm}
\caption{(color online).  (a) Energy level diagram showing charging energy, $U$, width of single particle energy level, $\Gamma$, depth of energy level, $\epsilon_0$, below the equilibrium Fermi level, $\mu$, and the co-tunneling exchange process leading to Kondo resonance resulting from (b) an extra peak in the density of states at the equilibrium Fermi level.  (c) Schematic of measurement setup.  (d) $G(V)$ \textit{vs.} $V$ for sample  \textbf{A} at $T/T_{\mathrm K} \lesssim 1.5$.  (e) $dI/dV$ \textit{vs.} $T$ also for sample \textbf{A}.  Solid black lines are fit to Eq. (\ref{GEK}) with $G_b=0$ (top), $G_b$ equal to optimum value as determined in text (middle), and $G_b$ determined through procedure used in Ref \cite{Scott2009} (bottom).  Gray dashed line is a fit to the analogous equation for an underscreened spin-1 Kondo effect (See Supplementary Information).}
\label{figure1}
\end{figure}

\begin{figure}[t!]
\begin{center}$
\begin{array}{cc}
\includegraphics[width = 2.7in]{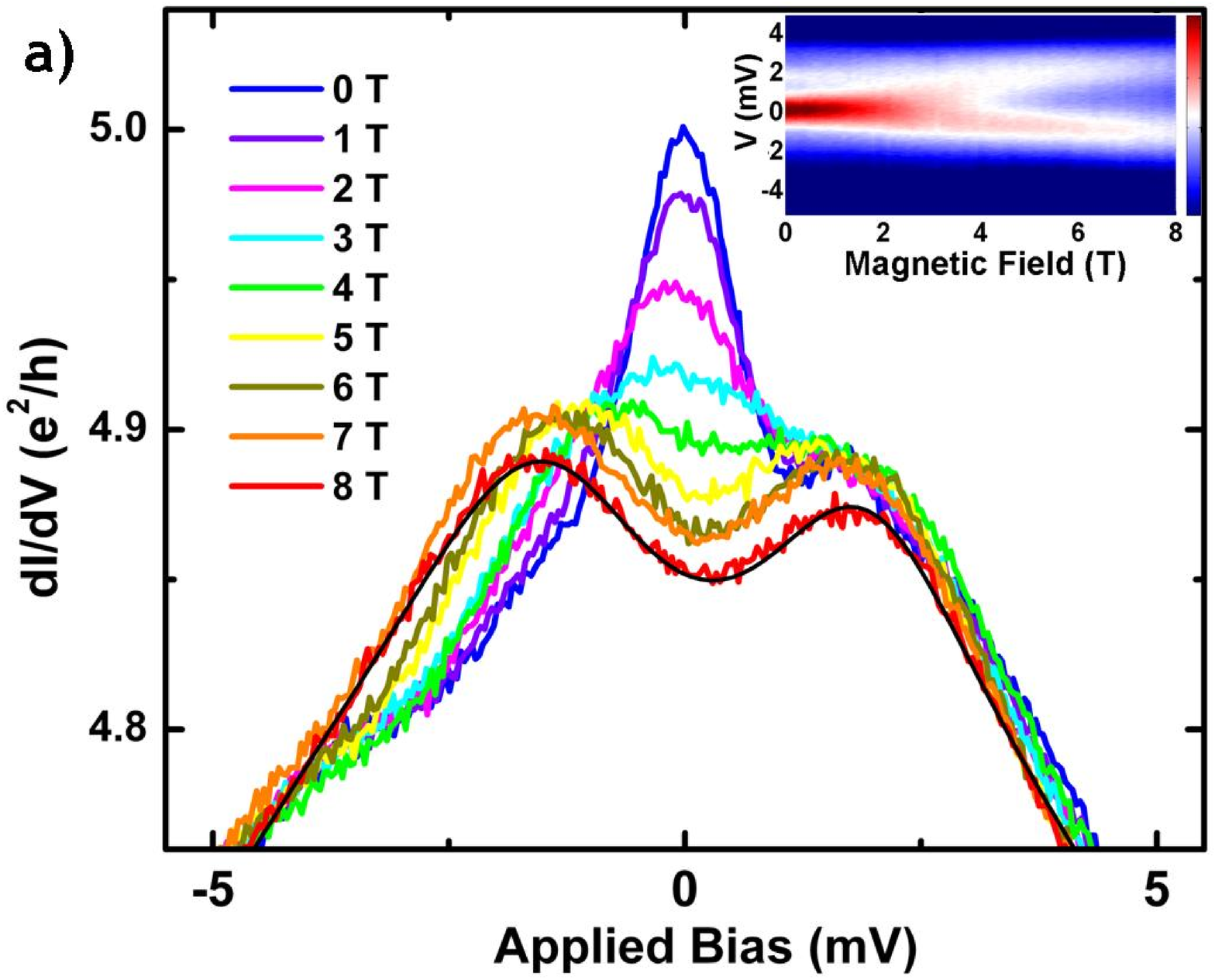}\\
\includegraphics[width = 2.7in]{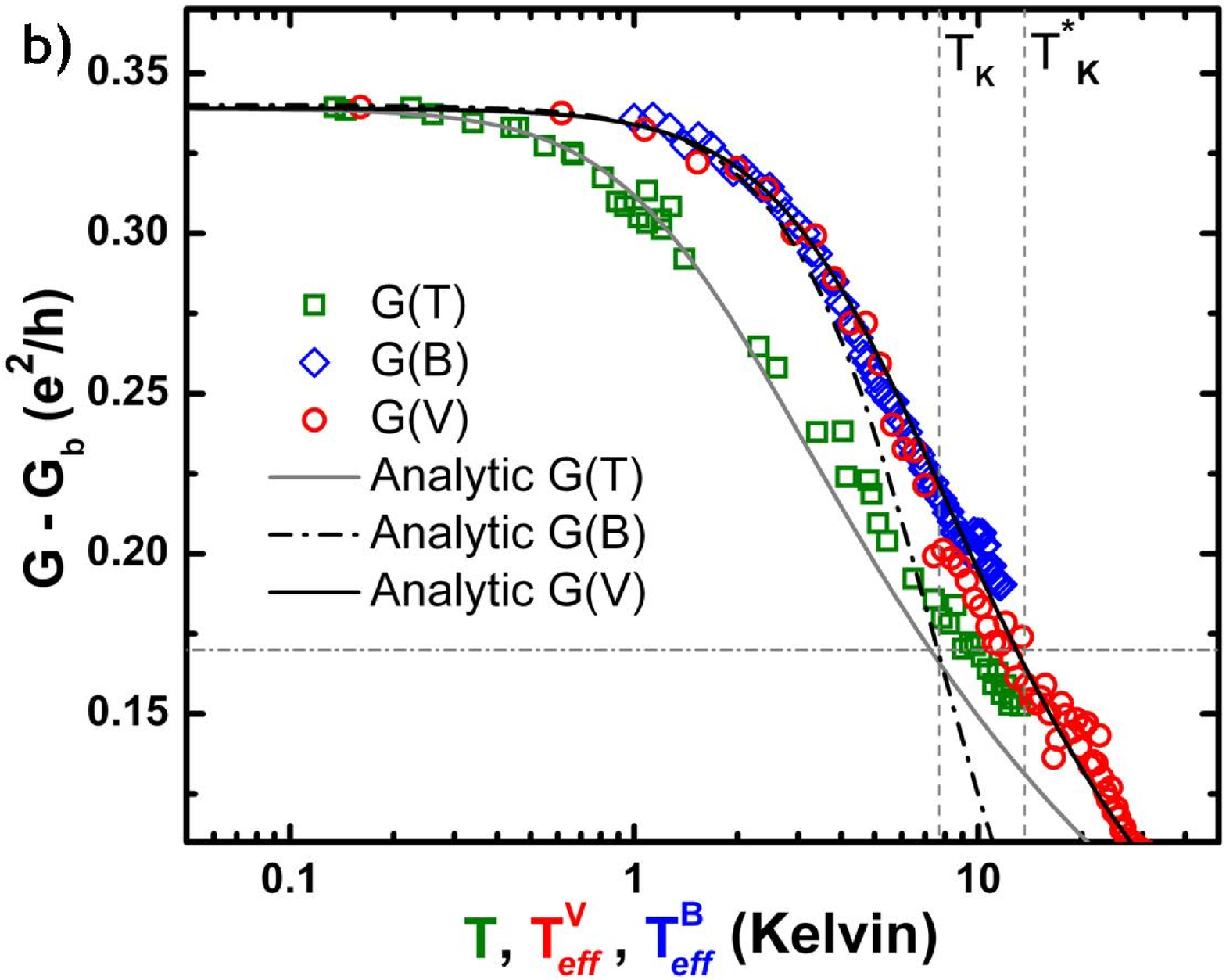}
\end{array}$
\end{center}
\vspace{-7mm}
\caption{(color online). (a) Conductance in sample \textbf{A} \textit{vs.} bias for several value of external magnetic field.  Black line represents a fit to data using an equation consisting of a sum of two Fano peaks plus a cubic background (see Suppl. Info).  Inset: Traces of $dI/dV \textit{vs.} V$ acquired at many values of $B$.  Colorbar indicates magnitude of $dI/dV$ ranging from 4.8$e^2/h$ (dark blue) to 5.0$e^2/h$ (dark red).  (b) Equilibrium conductance as a function of temperature (green squares); zero-bias, base temperature conductance as a function of effective temperature $T^B_{eff}$ (blue diamonds); and zero-field, base temperature conductance as a function of effective temperature $T^V_{eff}$ (red circles). The same $G_b$ is subtracted from all data shown.  Gray line is a fit to $G(T)$ data using Eq. (\ref{GEK}).  Black dash-dot line is a fit to $G(B)$ data using Eq. (\ref{GB1}).  Black line is a fit to $G(V)$ data using Eq. (\ref{GV1}).  The same value of $T_K$ is used for all three fits.  Horizontal gray dash-dot line signifies $\frac{1}{2}G_0$.}
\label{figure2}
\end{figure}

We have measured several samples that exhibit Kondo conductance with total $G$ above the unitary limit of $2e^2/h$.  The zero-bias peak in sample \textbf{A} (Figs.\ref{figure1}d,e and \ref{figure2}a,b), for example, exhibits typical Kondo resonance behavior.  However, the total conductance cannot result entirely from cotunneling via the same single particle level.  This implication of $G_b > 0$ is not surprising in the case of break junctions, where the interelectrode gap is frequently $< 1~nm$, and the tunneling current is exponentially sensitive to the electrode-molecule displacement.  Thus a likely source of $G_b$ is direct metal-to-metal tunneling between electrodes, or in some cases transport through a parallel portion of the junction that is not entirely broken.  The absence of distinctly asymmetric Fano lineshapes in the zero bias resonance peaks, as found in some break junction~\cite{Calvo2009} and STM experiments,~\cite{Nagaoka2002,Madhavan1998} suggests that $G_b$ originates from a non-resonant transport channel that does not interfere strongly with the cotunneling process responsible for the Kondo effect.  Therefore the resonant and non-resonant channels are not strongly coupled.  Possible origins of $G_b$ are discussed further in the Supplementary Information.

It has been proposed that data may be treated without accounting for extra conduction channels.\cite{ParksThesis,Grobis2008}  This can lead to coefficients closer to theoretically proposed values, but it also results in poor fits to data (Fig. \ref{figure1}e top) and erroneous estimations of $T_K$ for high conductance samples inconsistent with the commonly used approximation that the resonance peak's full width at half maximum is roughly $2\sqrt{2}k_{\mathrm{B}}T_K$~\cite{Nagaoka2002}.  In Ref.~\cite{Scott2009} similar devices were studied and a background contribution to the overall conductance, linear in applied bias, was subtracted from conductance measurements.  It was established via a fit to the trace $dI/dV$ $\textit{vs.}$ $V$, using a range of applied bias, $\pm{V}$, which incorporated both minima adjacent to the zero-bias resonance.  As an example, this ``full'' background subtraction is shown at the bottom of Fig. \ref{figure1}e when applied to sample \textbf{A}.

Rather than relying solely on an estimate for $T_K$ from Eq. (\ref{GEK}), we additionally utilize expressions for $G(V)$ and $G(B)$ while using a conductance offset, $G_b$, as a fitting parameter.\cite{Parks2007,Roch2009} In Ref.~\cite{Pletyukhov2012} an expression for the zero temperature conductance out of equilibrium, $G(V,0,0)$ for the spin-$\frac{1}{2}$ Kondo model was proposed.  This analytical form provides a description of the Kondo resonance peak analogous to Eq. (\ref{GEK}) in which
\begin{equation}
G(V) = G_0\left[1 + \frac{(2^{1/{s_1}}-1)x^2}{1 + b(|x|^{s_2}-1)}\right]^{-s_1},
\label{GV1}
\end{equation}
where $x = V/T^{\ast}_K$ and $({T^*_K}/T_K)^2 \approx \pi$.

The behavior of the equilibrium conductance as a function of magnetic field can also be used to characterize the device.  In Ref.~\cite{Munoz2013} an approach to transport in the particle-hole asymmetric single impurity Anderson molecule is developed based in terms of dual fermions.  When a finite magnetic field is included, this perturbative method yields a systematic expansion of the conductance $G(T,V,B)$.  For the spin-$\frac{1}{2}$ Kondo model, one finds~\cite{Merker2013}
\begin{equation}
G(B) \simeq \frac{e^2}{\hbar}\frac{8\xi}{3}\frac{1}{1+\left(\frac{\pi |g|\mu_BB}{4 k_{\mathrm{B}}T_K}\right)^2},
\label{GB1}
\end{equation}
where $\xi=3\Gamma_L \Gamma_R/(\Gamma_R+\Gamma_L)^2$ is a measure of the asymmetry in the dot-lead couplings
$\Gamma_L$ and $\Gamma_R$.  Bias and magnetic field are converted to units of effective temperature, where $k_{\mathrm{B}}T^V_{eff} \equiv eV + k_{\mathrm{B}}T_{base}$ and $k_{\mathrm{B}}T^B_{eff} \equiv |g|\mu_BB + k_{\mathrm{B}}T_{base}$, allowing for a quantitative comparison between the response to the different perturbations.  (The standard value of 2.0 was used in the conversion to units of $T^B_{eff}$.)

Adding the parameter $G_b$ to Eqs. (\ref{GV1}) and (\ref{GB1}) allows for fits to data of $G(T)$, $G(V)$, and $G(B)$. The confluence of these three relations leads to a value of $G_b$ that most nearly produces a single energy scale, $T_K$, for the three fits corresponding to a particular device.  Thus the optimum $G_b$ produces the best \textit{simultaneous} fits for a single $T_K$.  Certainly the parameters $G_0$ and $T_K$ will depend upon the magnitude of the parameter $G_b$; thus, the value of $G_b$ will also impact the values of the transport coefficients.  In Fig. \ref{figure2}b $G(T)$, $G(V)$, and $G(B)$ from sample \textbf{A} are plotted, along with their respective phenomenological curves, using $T_K = 7.32 K$ and $G_b = 4.7e^2/h$.

\begin{figure}[t!]
\begin{center}$
\begin{array}{cc}
\includegraphics[width = 1.6in]{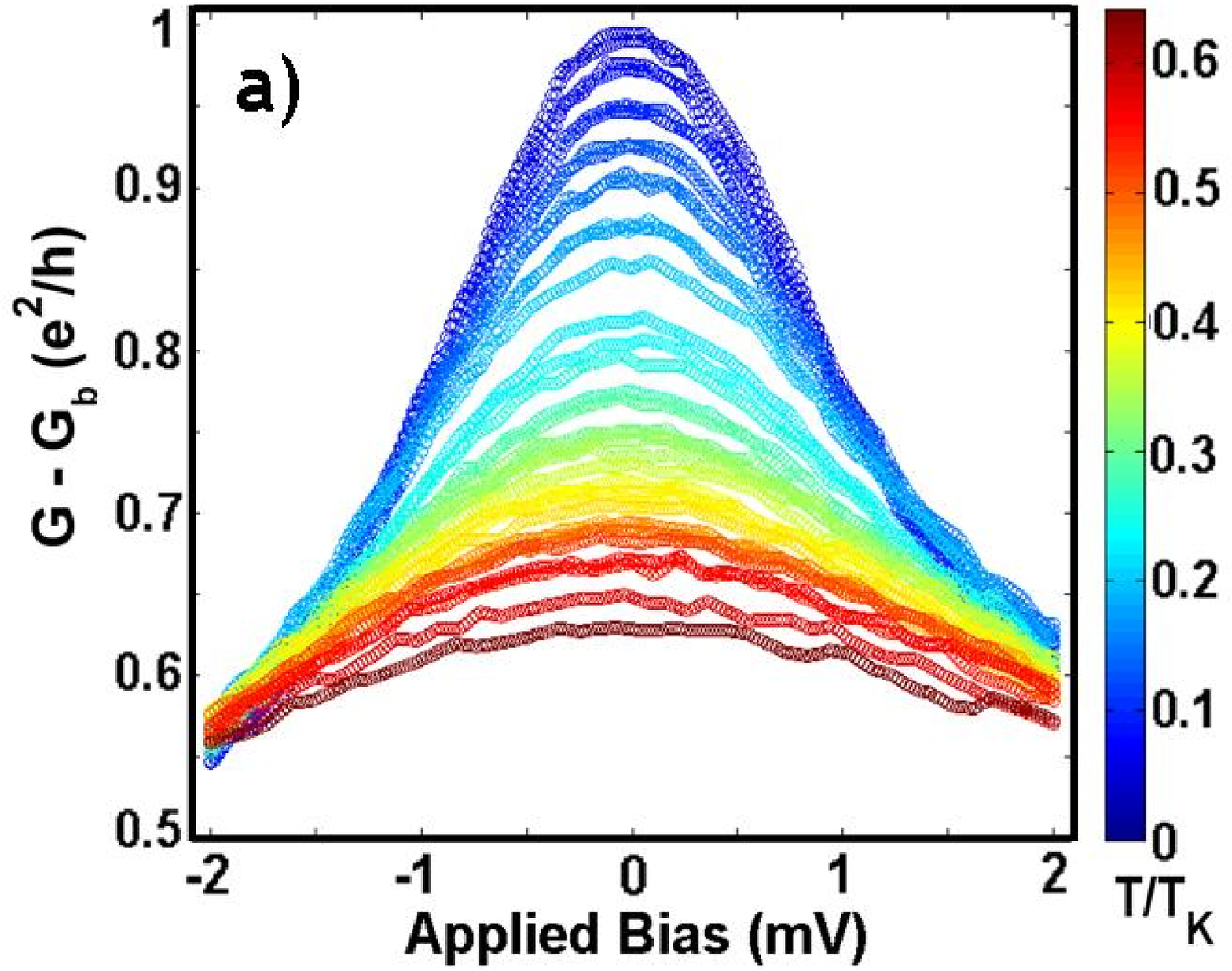} &
\includegraphics[width = 1.59in]{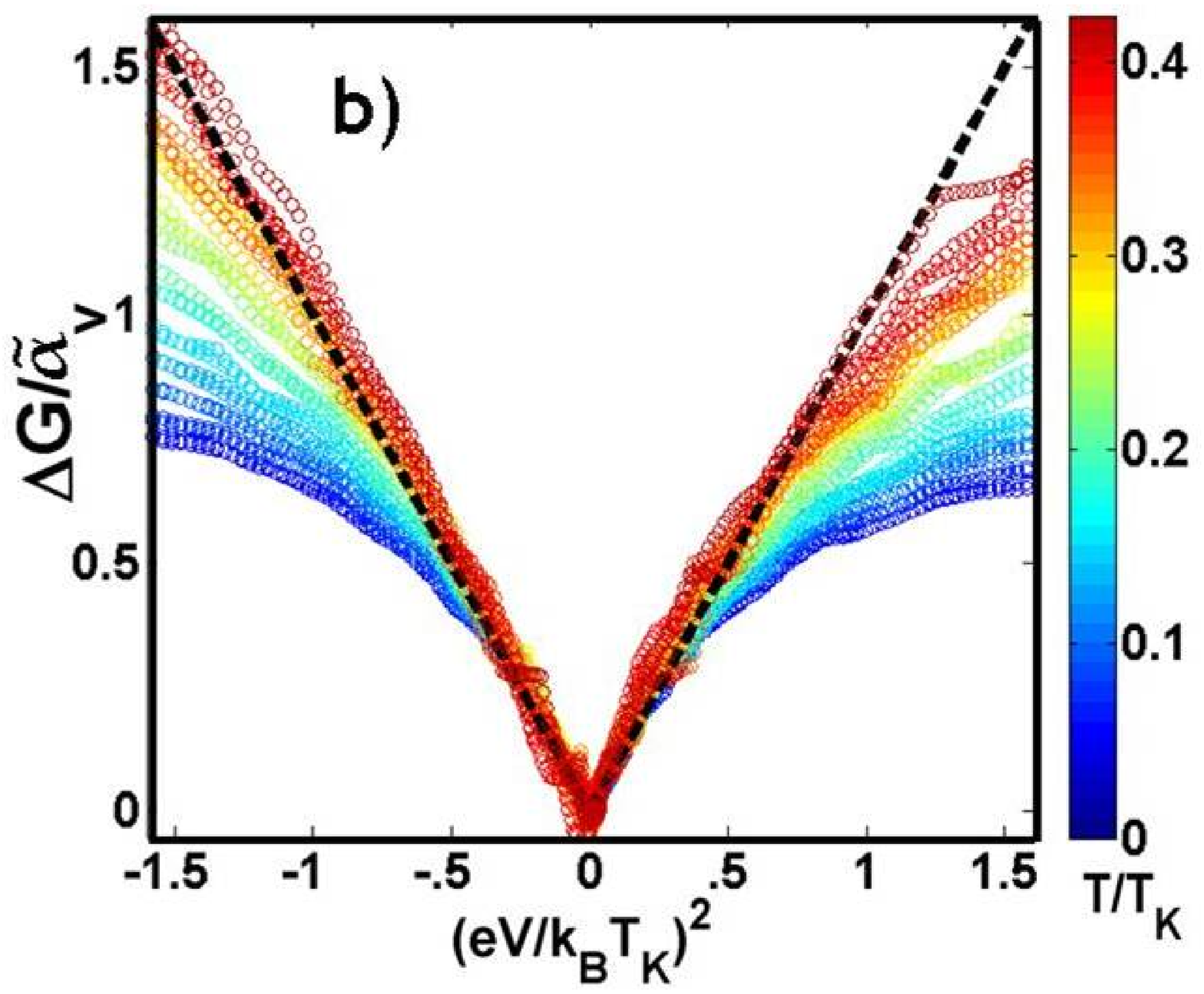} \\
\includegraphics[width=1.6in]{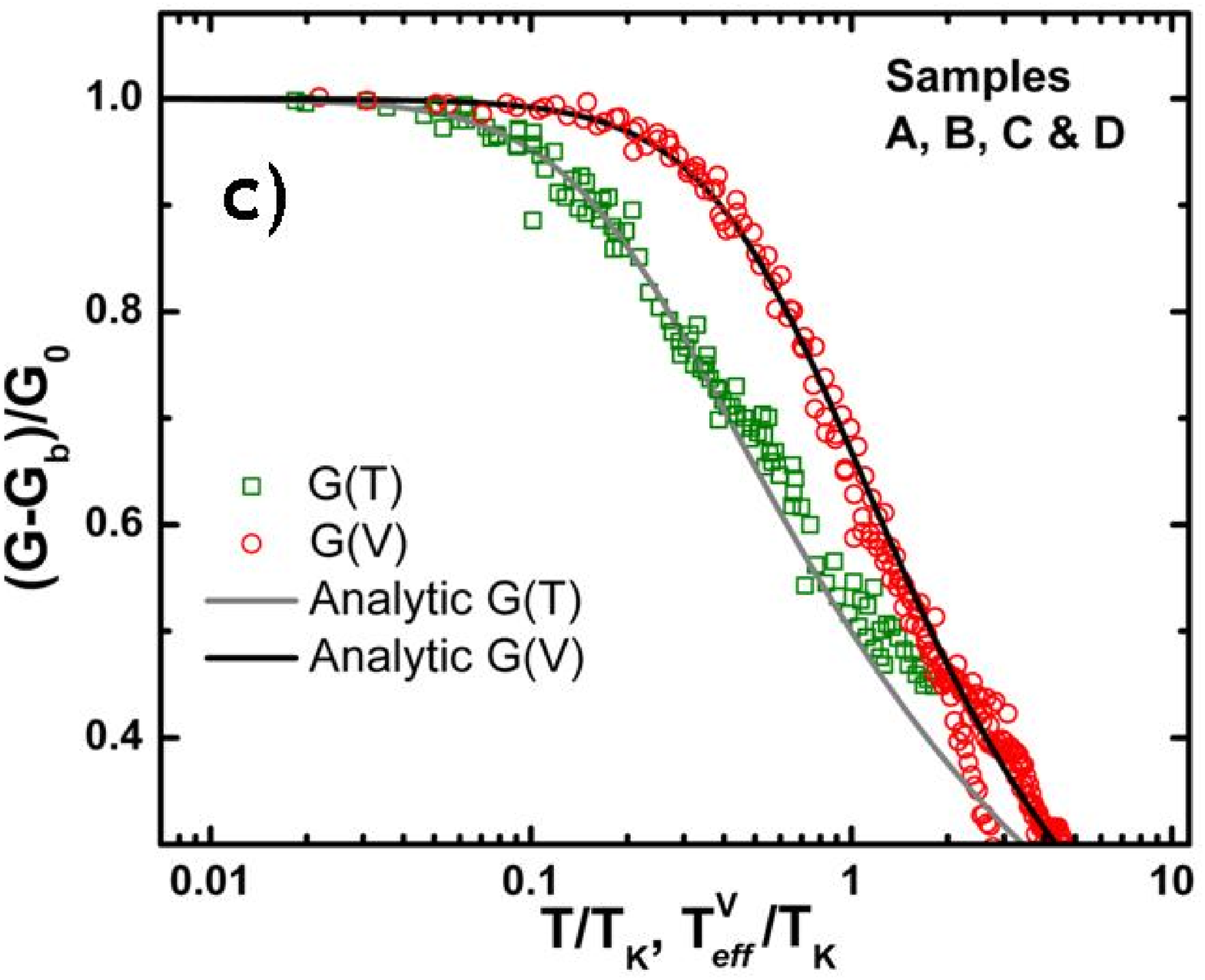} &
\includegraphics[width=1.6in]{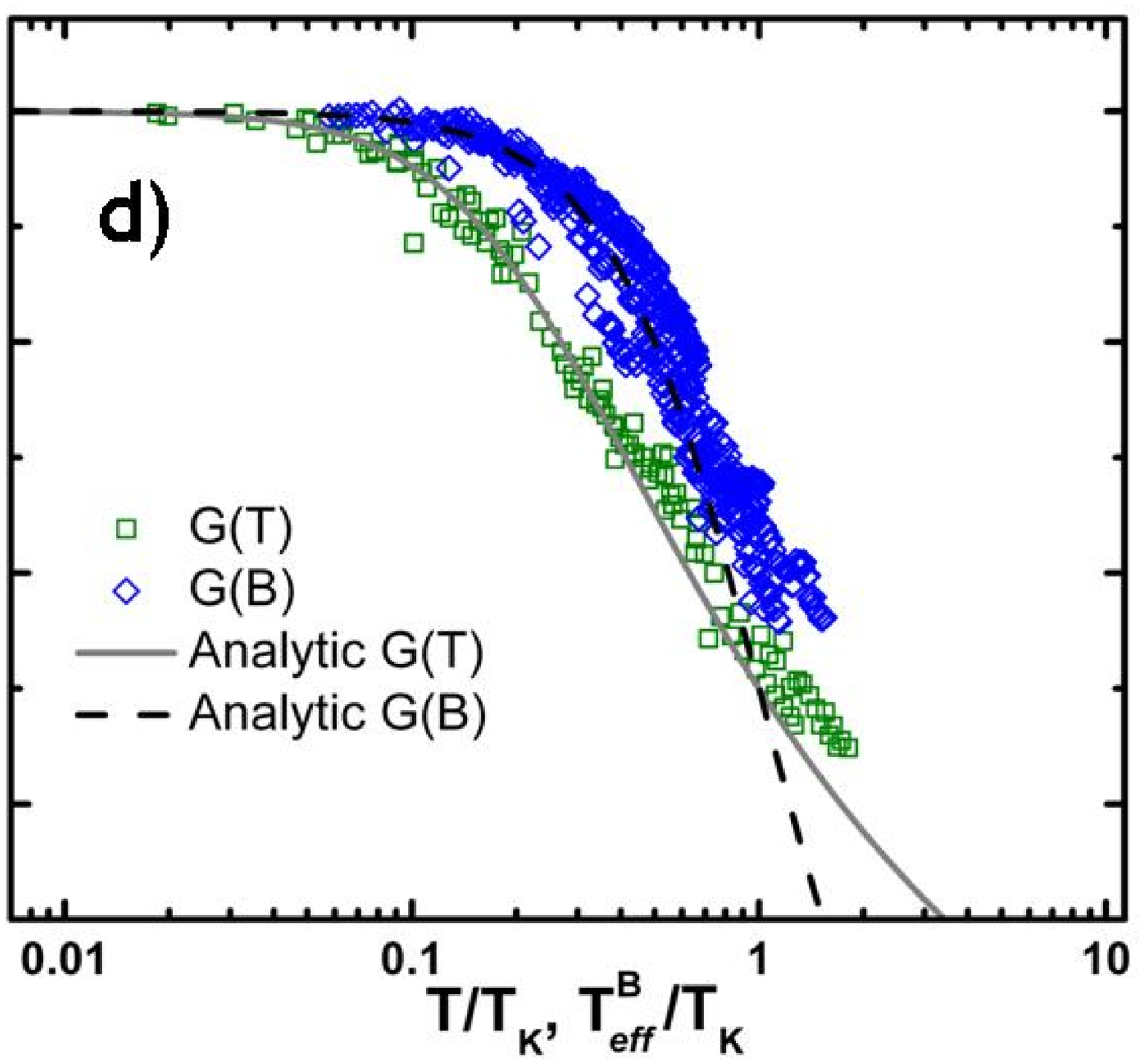}
\end{array}$
\end{center}
\vspace{-7mm}
\caption{(color online). (a) Conductance as a function of $V$ for sample \textbf{B} at $T/T_{\mathrm K} \lesssim .65$.  (b) Scaled conductance, $\Delta{G}/{\tilde{\alpha}_V}$, \textit{vs.} $(eV/k_{B}T_{K})^2$ for the same device at several temperatures, where $\Delta{G} = (1-G(T,V)/G(T,0))$, ${\tilde\alpha}_V \equiv c_T\alpha/(1 + c_T\left(\frac{\gamma}{\alpha} - 1\right)({T}/T_{\mathrm K})^2)$.  We used $\alpha = .11$ and $\gamma = .67$.  The dashed black line represents the associated universal curve.  The data conforms well to the scaling function below $(eV/k_{\mathrm B}T_{\mathrm K})^2 \lesssim .45$ for all temperatures plotted.  (c),(d) Normalized conductance plotted for 4 different samples, each possessing a different $T_{\mathrm K}$ between $\sim 7 - 31 K$.  $G(V)$ and $G(B)$ data are plotted separately for clarity.  Equilibrium conductance as a function of temperature (green squares) and (c) zero-field, base temperature conductance as a function of effective temperature $T^V_{eff}$ (red circles), or (d) zero-bias, base temperature conductance as a function of effective temperature $T^B_{eff}$ (blue diamonds).  Same \textit{y}-axis for (c) and (d).}
\label{figure3}
\end{figure}

Analysis of 9 samples exhibiting a single channel spin-$\frac{1}{2}$ Kondo effect, each with $7.32K\leq{T_{\mathrm K}}\leq31K$, resulted in the following average values: $c_T = 4.97 \pm 0.9$; $c_V = 0.74 \pm 0.14$; $c_B = 0.89 \pm 0.16$ found via fits to Eqs.~(\ref{GEK})-(\ref{GB1}) for energies up to $T/T_K < 0.1$.  These values may be compared to the theoretical predictions of $c_T = 5.38, c_V = 0.82$, and $c_B = 0.55$, in accord with the definition of $T_K$.  We also find that the effect of magnetic field is well described by the theoretically predicted curve for $T/T_K \lesssim 0.1$ (see Fig.~\ref{figure3}).  We stress that Eqs.~(\ref{GEK})-(\ref{GB1}) are strictly valid only for the particle-hole symmetric spin-$\frac{1}{2}$ Kondo model. To check for the internal consistency of our analysis, we compare with the method of Ref.~\cite{Munoz2013} generalized to include finite magnetic field.  From fitting the data of a representative sample to Eqs.(1)-(3), we find $\bar{c}_{T} = 4.91$, $\bar{c}_{V} = 0.72$, $\bar{c}_{B} = 1.49$ and $G(0,0,0)/G_{0} = 0.35$.  The large discrepancy between the two values for $c_B$ suggest that the $g$-factor of the device deviates from the assumed value $g=2.0$.  The renormalized parameters of the underlying Anderson model, as defined in Ref.~\cite{Munoz2013}, turned out to be $\tilde{u}=0.68$, $\tilde{\epsilon}_{d}=0.95$ and $\xi = 0.263$. While the value $\tilde{\epsilon}_{d}=0.95$ points to a large degree of particle-hole asymmetry, we note that according to Fig.2a of Ref.~\cite{Munoz2013}, $c_T$ at $\tilde{u}=0.68$ is hardly affected by it.

Derivations of the finite bias conductance behavior at non-zero temperature have been worked out in the context of both the particle-hole symmetric Anderson single level impurity model\cite{Konik2002,Oguri2005,Rincon2009a,Sela2009} and the Kondo model\cite{Schiller1995,Majumdar1998,Pustilnik2004,Doyon2006}.  A scaling function for the conductance out of equilibrium was outlined in Ref \cite{Grobis2008} based on a low energy expansion of a form suggested by Ref \cite{Nagaoka2002},
\begin{equation}
\footnotesize{G(T,V) = G(T,0)\left(1 - \frac{c_T\alpha}{1 + c_T\left(\frac{\gamma}{\alpha} - 1\right)
\left(\frac{T}{T_{\mathrm K}}\right)^2}\left(\frac{eV}{k_{\mathrm B}T_{\mathrm K}}\right)^2\right)},
\label{GTV1}
\end{equation}
\normalsize
where the coefficients $\alpha$ and $\gamma$ represent the zero-bias curvature of the Kondo resonance peak as a function of $V$ and the rate at which the resonance peak broadens and decays in amplitude as a function of $T$, respectively.  Using a value for $T_K$ inferred from Eq. (\ref{GEK}), some groups probing the Kondo state in nanotubes and quantum dots estimated $0.1 < \alpha < 0.25$ and $0.5 < \gamma < 1.65$,\cite{Grobis2008,ParksThesis,Kretinin2011}.  Analysis of our new samples find $\alpha = 0.13 \pm 0.018$ and $\gamma = 0.89 \pm 0.22$. Note, however, that $\gamma$ depends on the lead-to-dot coupling asymmetry $\xi$ even in the particle-hole symmetric spin-$\frac{1}{2}$ Kondo model.~\cite{Munoz2013}

In a past experiment measuring conductance in Ti/Au break junctions using two types of molecules two of the authors found values of $\alpha \approx 0.051$ and $\gamma \approx 0.107$.\cite{Scott2009}  A number of physical influences were postulated as potentially affecting the devices under study leading to the exceptionally small coefficient values. For instance electron-phonon coupling could lead to a renormalization of estimated parameters.\cite{Cornaglia2007,Elste2008}  Indeed, more recent theoretical studies have explored the impact of non-ideal conditions like finite on-site repulsion, tunneling barrier (\textit{i.e.} molecule-lead coupling) asymmetry, particle-hole asymmetry, and charge fluctuations related to the possible proximity to the mixed-valence regime.\cite{Sela2009,Rincon2009a,Roura-Bas2010,Munoz2013}  They generally conclude that these conditions may impact $\alpha$ by lowering its value in the range of $\sim 0.1 - 0.15$.  In the case of Ref. \cite{Sela2009} the authors anticipate $0.075 \lesssim \alpha \lesssim 0.3$.  Employing a limited version of the treatment described here (field response is not available for the previously recorded data) to the data of Ref. (\cite{Scott2009}) in order to account for background conductance, leads to the following coefficient values: $\alpha = 0.104 \pm 0.01; \gamma = 0.703 \pm 0.12; c_T = 5.25 \pm 1.0; c_V = 0.69 \pm 0.18$, which are in closer agreement with previously published results.  Note that while $c_T$ is evidently fixed by the choice of $s=0.21$ in Eq.~(\ref{GEK}) and it is anticipated that $c_V = \alpha{c_T}$, the values found in our experiments differ slightly.  This is due in part to the fact that $T_K$ is not established by Eq.~(\ref{GEK}) alone, and in part because of the more restrictive fitting range for the $c$'s.  For $\alpha$ and $\gamma$ the fits were limited to $T/T_K, eV/k_{\mathrm{B}}T_K, |g|\mu_BB/k_{\mathrm{B}}T_K < 0.2$ to $0.3$, depending on the sample.

We have measured conductance in break junction devices in the low energy equilibrium and non-equilibrium Kondo regime.  In some cases the peak conductance exceeds the unitary limit, highlighting the need for parallel transport channels to be taken into consideration.  Kondo transport out of equilibrium remains a difficult problem.  Although there are solutions that are asymptotically correct in some regions, we have found that a consistent picture may be obtained by bringing together various approaches.  The use of multiple analytical expressions for the dependence on temperature, bias, and magnetic field provides a more reliable means of extracting the emergent energy scale, $T_K$, and it enables us to converge upon the proper conductance offset, which allows for the separation of background conductance from the cotunneling exchange process associated with the Kondo effect.  We have applied this method to a group of new samples and found transport coefficients consistent with previous experiments and with theoretical expectations, particularly when non-ideal conditions are considered.  When a limited version of this technique is applied to older data it also brings coefficient values in-line with expectations. Further inconsistencies may be addressed with  additional considerations, for example accounting for a background conduction with a weak temperature or bias dependence.

GDS would like to thank R. Willett and M. Peabody for technical support, and T.-C. Hu and R. Kopf for assistance with sample fabrication.  DN acknowledges NSF DMR-0855607. SK and EM would like to thank T. Costi, and acknowledge support by the Comisi\'{o}n Nacional de Investigaci\'{o}n Cient\'{i}fica y Tecnol\'{o}gica (CONICYT), grant No. 11100064 and the German Academic Exchange Service (DAAD) under grant No. 52636698.

\newpage
\
\newpage

\section{Supplementary Information}

\subsection{Non-spin-$\frac{1}{2}$ Kondo Effect}

It is possible to find C$_{60}$ with spin $S = 1$ in break junction devices,\cite{Roch2009,Parks2010} thus consideration was taken in this experiment to identify the spin state of the impurity in order to ensure that the analysis was applied only to samples containing a magnetic moment with $S = \frac{1}{2}$.  The equilibrium temperature dependence of the Kondo resonance described by Eq. (4) pertains only to the case of a spin-$\frac{1}{2}$ impurity coupled to a single conduction channel.  More generally, the following phenomenological function, connecting the low ($T \ll T_K$) and intermediate energy ($T \sim T_K$) regimes, can be used to describe the characteristic response of the system with respect to temperature,
\begin{equation}
G(T) = G(0)\left[1 + \left(\frac{T}{T'_K}\right)^{\xi_S} \right]^{-\alpha_S}.
\label{GTgeneral}
\end{equation}
Here $T'_K = T_K/(2^{1/{\alpha_S}} - 1)^{1/{\xi_S}}$ while $\alpha_S$ and $\xi_S$ are constants that depend upon the spin of the magnetic impurity and the number of conduction channels that may participate in screening the impurity.  For an impurity with spin state $\frac{1}{2}$ coupled to one channel we use $\alpha_S = 0.21$ and $\xi_S = 2$.

To distinguish between a spin-$\frac{1}{2}$ fully screened Kondo effect and a spin-1 underscreened (or fully screened) Kondo effect we attempted to fit the respective versions of the empirical form to the equilibrium data.  A system is said to be fully screened when $2S = n$, where $S$ is the spin state and $n$ is the number of screening channels.  If $2S < n$ then the system is said to be underscreened.  Because our devices effectively possess only one screening channel, we were primarily interested in the possibility of an underscreened effect.\cite{Potok2007}  The $G(T)$ data was fit to Eq. (\ref{GTgeneral}) using $\alpha_S = 0.33$ and $\xi_S = 0.89$.  These values were found via fits to the numerical renormalization group analysis in Ref. \cite{Parks2010}.  In most cases this produced very poor fits compared to the equation for the spin-$\frac{1}{2}$ impurity, as shown in Fig. 1e.  For the sample in Fig. \ref{figure1} the fit of $G(T)$ for $S = \frac{1}{2}$ and $S = 1$ produced roughly equal quality fits.  Additionally, the magnetic field dependence of the base temperature conductance did not follow the expected format associated with a spin-$\frac{1}{2}$ impurity.  Due to the ambiguous nature of this sample, and two others with similar properties, it was not included in the scaling analysis discussed in the text.

\begin{figure}[t!]
\begin{center}
\includegraphics[scale = .65]{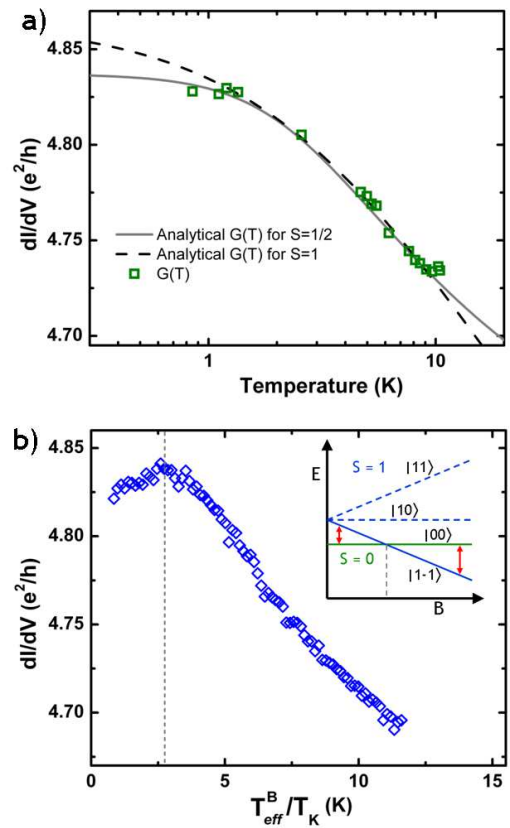}
\end{center}
\vspace{-7mm}
\caption{(a) $G(T,0,0)$ \textit{vs.} $T$ for a sample that does not exhibit clear spin-$\frac{1}{2}$ characteristics.  Gray line is a fit to the empirical expression in Eq. (2).  The black dashed line is a fit to the analogous equation for an underscreened spin-1 Kondo effect shown in Eq. (S1).  The quality of the fits is similar.  (b) $G(0,0,B)$ \textit{vs.} $B$ for the same sample.  Of particular note is the increase in zero bias conductance as the magnetic field is increased up to approximately $2.6~K$ in units of effective temperature.  While we were not able to resolve two separate peaks in traces of $dI/dV$ \textit{vs.} $V$ acquired on the left side of the vertical gray dashed line, this behavior is consistent with a spin-1 Kondo effect in which the magnetic field drives a transition between a singlet and a triplet ground state.  Inset: Energy as a function of magnetic field. Vertical gray dashed line corresponds to a magnetic field value that results in a degeneracy between the singlet $\ket{0~0}$ state and the triplet $\ket{1~{-1}}$ state.}
\label{figure1}
\end{figure}

\subsection{Origins of $G_b$}

The existence of parallel transport in the Kondo regime of nanoscale devices is neither new nor exceptional.  Electrostatically defined semiconductor quantum dots, for example, in the Kondo regime can exhibit co-tunneling that is not Kondo-related, but its contribution to the total conductance is typically quite small in these structures.  Metallic break junctions have been used with great success to investigate many of the same transport properties observed in semiconductor quantum dots. One notable distinction between these structures is the origin of parallel transport and its corresponding magnitude.  Much like an STM experiment performed with a relatively blunt tip, transport in break junctions is not necessarily limited to a single atomic channel.\cite{Scheer2000}  Rather there may be more than one channel contributing to the measured conductivity since tunneling current is highly dependent on the detailed morphology of the tip (or junction, in our case).

While there may be more than one transport pathway available, it is our understanding that the zero bias resonance peaks result from the interactions between conduction electrons and an unpaired spin on a molecule.  The idea that the local moment in our devices is formed within the metal itself is doubtful.  Reference \cite{Calvo2009} describes measurements of atomic contacts in which the single channel Kondo effect can be seen in clean devices (\emph{i.e.} without molecules) that are not completely broken.  The argument put forth by the authors of Ref. \cite{Calvo2009} relied on their use of ferromagnetic metals that contain strongly localized \emph{d}-orbitals.  The metal of our break junctions is not ferromagnetic, so their premise will not apply equally to our system.  The general lack of zero bias resonance peaks in our control samples additionally signifies that the effect is at least associated with the molecules, which may provide the local moment required for Kondo transport.

The absence of any strongly asymmetric Fano-type lineshapes in our observed zero-bias resonances indicates that if non-resonant parallel transport is occurring then it is not strongly interfering with the resonant tunneling via the local moment.\cite{Schiller2000,Jamneala2000,Nagaoka2002,Calvo2012}  Therefore it is doubtful that the molecule is providing both a path for the co-tunneling exchange process leading to the Kondo effect and a bridge for the parallel channels as this would likely lead to interference between resonant and non-resonant processes.

This argument implies the dominant source of $G_b$ is metal to metal tunneling at a point in the junction apart from the molecule.  The possibility also exists that $G_b$ could be due to transport through an unbroken channel within the junction.  We note that $G_b$ may be greater than zero whether or not $G(0,0,0) > 2e^2/h$.  This proposed system may entail a scenario analogous to that described in Ref. \cite{Sato2005}, in which transport is measured through an electrostatically defined semiconductor quantum wire that is side-coupled to a similarly defined quantum dot.  Changes in the coupling strength between the quantum wire and dot led to measured changes in conductance of the wire together with an evolving Fano-Kondo resonance.  Our devices may have a comparable organization, but with the equivalent of a very weak coupling between the dot and the wire.

We maintain that the measured conductance near zero bias originates from resonant transport through a single molecule in addition to non-interfering, non-resonant transport through a parallel channel apart from the molecule.  As long as the coupling between the different channels is small, the application of our method is equally valid whether the conductance attributable to $G_b$ is from tunneling or direct contact at another site.  Since the parallel transport in either situation not significantly dependent on $T$, $V$, or $B$ at the energy scale we probe, it is still reasonable to model it as a constant, and thus the proposed method of extracting $T_K$ can (and should) be applied.

\subsection{Data Fitting}

Equations 4 through 7, describing the conductance response to different perturbations, are valid only up to at most intermediate energies ($T/T_K, eV/k_BT_K, |g|\mu_BB/k_BT_K \sim 1$).  As the perturbations are further increased relative to $T_K$, it is expected that the measured conductance will deviate from these expressions due to the onset of transport processes unrelated to Kondo-assisted tunneling, yet this transition is not expected to show any clear demarcation.  In the case of finite magnetic field (Eq. (7)), it is expected that the conductance data will exhibit a stronger deviation from the analytical formula since the theory is perturbative in the field, as opposed to non-perturbative results derived from numerical renormalization group techniques.

To establish a consistent means of determining the low energy zero bias conductance for all traces of $dI/dV \textit{vs.} V$ a parabola is fit to the central portion of the resonance.  It's maximum value is taken as the zero bias conductance.  For traces in which the resonance peak is found to be offset from $V = 0~mV$, commonly by $< \pm{0.5~mV}$ due to amplifier drift, the peak is re-centered for the ensuing analysis.

The uncertainty in the coefficient values we found, as well as their corresponding values of $G_0$ and $T_K$, arise from a few different sources of error.  Measurements of $dI/dV$ at each $V$ data point included some random noise, which varied in magnitude from device to device.  There is some uncertainty in the thermometry calibration and temperature stability during the course of acquiring a trace of $dI/dV \textit{vs.} V$.  Finally, the range of data points included in fits can impact the value of the fitting parameters.  This is especially noteworthy for conductance probed as a function of temperature.  The comparatively limited control and lower number of data points collected as a function of temperature, in contrast to what was acquired with respect to bias and magnetic field, makes the extraction of $T_K$ susceptible to significant differences for small changes in the fitting range.

To maintain the validity of our analysis the fit ranges were constrained to lower energies so as to avoid fitting to data outside the applicable range of the functions.  For a given sample the data range used for the fits was approximately equal for $G(T)$ and $G(V)$, and was commonly somewhat smaller for $G(B)$ due to the limited range of validity of Eq. (6), as mentioned above.  For a given sample small differences between the range of $G(T)$ and $G(V)$ occur because of the larger spacing between data points for the measurements of $G(T)$.  The quality of the fits was compared on the basis of the root mean square deviation of the data from the analytical expressions.  For $G(V)$ and $G(T)$, fits were performed up to $T/T_K$ and $T^V_{eff}/T_K \lesssim 0.2$ to $0.4$ depending on the sample, and for $G(B)$ fits were performed up to $T^B_{eff}/T_K \lesssim 0.08$ to $0.15$ where the numbers were also sample dependent.

In accord with the Zeeman effect, the application of an external magnetic field, $B$, lifts the spin degeneracy of the many-body Kondo singlet, and results in the splitting of the zero-bias resonance peak.  As $B$ is increased from $0~T$ to some field value $B_C$, the Kondo resonance, $dI/dV$ \textit{vs.} $V$, is observed to broaden and decrease in height.  For $B > B_C$ the splitting is large enough that two peaks can be resolved.  Their position and amplitude is ascertained from by a fit to the sum of two fano lineshapes plus a cubic background,\cite{Kretinin2011}
\begin{equation}
\frac{dI}{dV} = y_o + A_1\frac{(\epsilon_1 + q_1)^2}{1 + {\epsilon_1}^2} + A_2\frac{(\epsilon_2 + q_2)^2}{1 + {\epsilon_2}^2} + B|V|^3,
\label{fano2}
\end{equation}
where ${\epsilon}_{1,2} = (V - x_{1,2})/w_{1,2}$, $x_{1,2}$ is the center of a peak, and $w_{1,2}$ is its half-width at half maximum.  This is demonstrated in Fig. 2a, in which the solid black line is a fit of Eq. (\ref{fano2}) to the $dI/dV$ \textit{vs.} $V$ data corresponding to $B = 8 T$.

\subsection{Zero bias anomaly statistics}

Thermally-assisted electromigration was performed on 62 break junction samples.  Of the 35 samples decorated with C$_{60}$, 12 exhibited some sort of zero bias anomaly.  Clear signatures of the single channel fully screened spin-1/2 Kondo effect were displayed in 9 of these 12.  Molecules were not deposited on 27 samples.  These were utilized as control devices in order to test for the likelihood of observing zero bias anomalies unrelated to transport through a C$_{60}$ molecule.  In this control group a zero bias conductance peak was observed in two devices, at least one of which did not exhibit typical Kondo transport behavior.  The rate with which we expect to see Kondo-related transport are commensurate with the findings of past investigations.\cite{Liang2002,Yu2004,Scott2009}

The rate of occurrence of a zero bias conductance peak in samples decorated with C$_{60}$ is much greater than that of the control devices.  This leads us to believe that transport in these samples is related to the C$_{60}$.  However, we acknowledge the possibility that transport leading to the resonance peak in the samples identified as $S = \frac{1}{2}$ Kondo effect due to a single unpaired spin situated on a C$_{60}$ molecules could be due to an alternate process or composition.

Another experimental realization of the Kondo effect has been demonstrated in chemically homogenous magnetic nanocontacts without any clearly defined impurity.  In Ref. \cite{Calvo2009} the authors argue that a small portion of one contact may function as the local moment.  This local moment could form in the contact atoms because of their smaller coordination and reduced symmetry, compared with the bulk.  The metal used in our break junctions is Pd.  While Pd is paramagnetic in its bulk state, only a small change in band structure is required to trigger itinerant ferromagnetism.  Nanoscale structuring via the electromigration technique will affect grain boundaries and lead to undercoordinated atoms at the break junction site, which can alter the density of states at the Fermi level thereby affecting the local magnetization.\cite{Lee1998,Sampedro2003,Scott2010a}

Our study is primarily focused on probing the Kondo state, regardless of what constitutes the local impurity.  Therefore, the possibility that transport may be occurring through a local moment other than a single C$_{60}$ molecule does not negate the consistency of the results.  Furthermore, part of this study is aimed at contrasting the results obtained from different experimental realizations of the Kondo effect, and a magnetic impurity in a break junction device, regardless of its makeup, constitutes a fundamentally different system compared to experiments probing Kondo scattering in semiconductor and nanotube structures.

\bibliographystyle{apsrev}

\bibliography{KondoScaling2012}

\end{document}